\documentclass[10pt,journal,compsoc]{IEEEtran}
\ifCLASSOPTIONcompsoc
  \usepackage[nocompress]{cite}
\else
  \usepackage{cite}
\fi
\ifCLASSINFOpdf
\else
\fi
\usepackage{array}
\usepackage{amsmath}
\usepackage{amsbsy}
\usepackage{amssymb}
\usepackage{algorithm}
\usepackage{algpseudocode}
\usepackage{float}
\usepackage{listings}
\usepackage{url}
\usepackage{hyperref}
\usepackage{graphicx}
\usepackage{enumitem}
\usepackage{tikz}
\usepackage{subcaption}
\usepackage{todonotes}

\usepackage{tikz-cd}

\graphicspath{{./img/}}

\definecolor{test}{rgb}{0.82, 0.1, 0.26}

\definecolor{mGreen}{rgb}{0,0.6,0}
\definecolor{mGray}{rgb}{0.5,0.5,0.5}
\definecolor{mPurple}{rgb}{0.58,0,0.82}
\definecolor{backgroundColour}{rgb}{0.95,0.95,0.92}

\lstdefinestyle{argvListing}
  {
  language=matlab,
  aboveskip=3mm,
  belowskip=3mm,
  showstringspaces=false,
  columns=flexible,
  basicstyle={\small\ttfamily},
  numbers=none,
  numberstyle=\tiny\color{gray},
  commentstyle=\color{dkgreen},
  stringstyle=\color{mauve},
  breaklines=true,
  breakatwhitespace=true,
  tabsize=3,
  inputencoding=ansinew,
  extendedchars=true,
  literate={æ}{{\ae}}1 {å}{{\aa}}1 {ø}{{\o}}1 {Æ}{{\AE}}1 {Å}{{\AA}}1 {Ø}{{\O}}1,
}
\lstdefinestyle{customasm}{
  backgroundcolor=\color{backgroundColour},   
  xleftmargin=\parindent,
  basicstyle=\footnotesize,
    tabsize=2,
    keepspaces=true,       
            breaklines=true,                 
      numbers=left, 
      xleftmargin=.05\textwidth,
    showtabs=false,                  
    showspaces=false,                
    breakatwhitespace=false,         
    showstringspaces=false,
    numbersep=5pt,                  
}

\begin{document}

%
\title{Stack-based Buffer Overflow Detection using Recurrent Neural Networks}
%
%
%
%

\author{William~Arild~Dahl,~
        L{\'a}szl{\'o}~Erd{\H o}di~
        and~Fabio~Massimo~Zennaro
\IEEEcompsocitemizethanks{\IEEEcompsocthanksitem W.A. Dahl, L. Erd{\H o}di and F.M. Zennaro are with the Department of Informatics, University of Oslo, 0316 Oslo, Norway\protect\\
E-mail: williada@ifi.uio.no, laszloe@ifi.uio.no, fabiomz@ifi.uio.no}
\thanks{Manuscript received: }}

%
%

\markboth{IEEE Transactions}
{Shell \MakeLowercase{\textit{et al.}}: Bare Demo of IEEEtran.cls for Computer Society Journals}
%



\IEEEtitleabstractindextext{%

\begin{abstract}
Detecting vulnerabilities in software is a critical challenge in the development and deployment of applications. One of the most known and dangerous vulnerabilities is stack-based buffer overflows, which may allow potential attackers to execute malicious code. In this paper we consider the use of modern machine learning models, specifically recurrent neural networks, to detect stack-based buffer overflow vulnerabilities in the assembly code of a program. Since assembly code is a generic and common representation, focusing on this language allows us to potentially consider programs written in several different programming languages. Moreover, we subscribe to the hypothesis that code may be treated as natural language, and thus we process assembly code using standard architectures commonly employed in natural language processing. We perform a set of experiments aimed at confirming the validity of the natural language hypothesis and the feasibility of using recurrent neural networks for detecting vulnerabilities. Our results show that our architecture is able to capture subtle stack-based buffer overflow vulnerabilities that strongly depend on the context, thus suggesting that this approach may be extended to real-world setting, as well as to other forms of vulnerability detection.
\end{abstract}

}

\maketitle

\IEEEdisplaynontitleabstractindextext

%
\IEEEpeerreviewmaketitle

\IEEEraisesectionheading{\section{Introduction}\label{sec:introduction}}

%
%
%
%



\IEEEPARstart{M}{odern} software applications deal with large quantities of data, including critical and sensitive data. The high demand for services has driven developers to create software at a fast pace, thus introducing flaws and bugs in the code. The number of software vulnerabilities discovered each year has steadily increased over the last decades, with a record of 16,665 reported discoveries in 2018 \cite{vulnYear}. 
Reducing the number of vulnerabilities and potential security breaches is a crucial challenge.

Exploiting flaws in code has become a multi-billion dollar a year business.
Malware developed to exploit code vulnerabilities is an ever-increasing problem for users, corporations, and governments worldwide. When the Shadow Brokers \cite{shadowBroke} leaked the EternalBlue exploit \cite{etertalBlue}, which gave rise to the WannaCry ransomware \cite{wannaCry}, ensuing damages have been estimated to be up to 8 billion USD globally.


An approach to reduce the number of vulnerabilities is based on the use of automatic tools for the identification and the patching of vulnerabilities. In this paper we explore the use of machine learning techniques for vulnerability detection. More precisely, we focus on the use of neural networks to identify the presence of potential stack-based buffer overflow vulnerabilities in assembly code. We make the assumption that code can be treated as a form of language, and we process it using recurrent neural networks (RNN) based on long short-term memory (LSTM) cells \cite{hochreiter1997long}. RNNs are a class of neural networks designed to handle input vectors of arbitrary lengths, and to extrapolate context from sequences. Such models were proven to be effective in tasks related to natural language processing (NLP) and they can be easily adapted to process code. 

To evaluate the effectiveness of RNNs in processing assembly files in order to spot buffer overflow vulnerabilities we carried out a set of empirical simulations. Our results, although preliminary, confirm the hypothesis that code may be dealt as a language, and they corroborate the intuition that neural network models may be successfully adapted to carry out vulnerability detection, in particular stack-based buffer overflow.




The rest of the paper is structured as follows. Section \ref{sec:Background} briefly introduces relevant concepts from information security and from machine learning. Section \ref{sec:RelatedWork} offers an overview of the state of the art in the application of machine learning to vulnerability detection. Section \ref{sec:ProblemStatement} precisely defines the problem we aim at tackling, as well as the assumption we will make in our study. Section \ref{sec:Methodology} provides details on our methodology, including the process of data generation, data pre-processing and model design. Section \ref{sec:Experiments} reports a set of experiments aimed at validating our hypothesis, and Section \ref{sec:Discussion} offers an overall evaluation of our results. Finally, Section \ref{sec:Conclusion} summarizes our results and suggests possible future directions of research.

\section{Background} \label{sec:Background}
In this section we provide a brief review of the main ideas from computer security (buffer overflow) and machine learning (RNNs) that are relevant to our work.

\subsection{Buffer overflow}
Vulnerabilities in software can have serious consequences from the point of view of computer security. For instance, without proper input data validation or accurate memory management malicious actors can modify the intended program flow of the software, thus leading to arbitrary code execution. The exploitation of a software vulnerability is frequently the starting point of complex attacks; for instance, the vulnerable software can execute a malicious payload to open a communication channel back to the attacker's computer, then download and execute a script from the attacker's location. Perfect software would be an ideal solution to protect systems against similar attacks, but such a possibility is not realistic; based on existing surveys, typical software vulnerabilities such as \emph{buffer overflows}, \emph{use after free}, \emph{double free vulnerabilities}, \emph{access control problems}, \emph{race conditions} or \emph{authentication weaknesses} appear frequently in software products. Despite many defensive measures were taken by the compilers and the operating systems, several new type of modern exploitation still appear time by time, e.g. \cite{wang2019layered}. A more feasible option to mitigate such risks is to discover software vulnerabilities as early as possible, for instance using fuzz testing or code analysis. In this work we focus on detecting stack-based buffer overflows. Finding abnormal program behavior such as buffer overflow errors is crucial in all computers and devices \cite{zhai2015method}.

An operating system executes binary code in the process virtual address space. In this space, the operating system keeps the executable code only of the binary, together with all the binary code that was linked (e.g., operating system APIs). In addition to the executable binary code, the virtual address space contains data sections such as the stack segments for each running thread, heap blocks and global variables. The partitioning of the memory space allows the operating system to keep processes separated, and it guarantees that different running binaries have no direct access to each other. The weak point of this data structuring is the fact that the code and the data are together in the same virtual address space; if there is no appropriate protection in place, this can lead to different types of memory corruption, such as executing data as code, or overwriting the code using data.

Buffer overflows compromise the virtual address space by overflowing the storage place that was allocated for the data. This action can lead to the modification of the virtual address space, for instance, by overwriting the heap chunks or the stack frames of the running methods inside the binary. In case of stack-based buffer overflows, the return pointer of a method that is under execution is modified by overrunning a local variable inside the method.
In some simple cases, stack-based buffer overflow may be carried out by exploiting methods that are vulnerable by default, such as the C methods \emph{gets}, \emph{puts}, \emph{strcpy}; these methods notoriously lack input validation, and, if the programmer does not perform proper size checking, this weakness can be easily used to successfully overflow the buffer. More complex and refined examples of stack-based buffer overflow may happen by exploiting a chain of several minor vulnerabilities, such as a chain of wrong input data that the attacker tries to set intentionally to overrun the vulnerable buffer. Our focus is mainly on the easiest case where the binary vulnerability is determined by the use of the aforementioned vulnerable methods, and where the solution corresponds to detecting the presence of such method calls without proper input checking.

\subsection{Machine learning}

The problem of detecting code that is potentially vulnerable to stack-based buffer overflow can be cast as a machine learning problem. In other words, we consider the possibility of training a model which could receive in input samples of code, and which could produce in output a signal denoting whether a potential buffer overflow vulnerability is present in the code. Neural networks have proven to be a powerful and flexible family of models to learn detection systems; for instance convolutional neural networks (CNNs) can be very effectively applied to image detection and recognition \cite{krizhevsky2012imagenet}.

Since our problem requires processing inputs of arbitrary length (i.e., code made up of a variable number of lines) and to evaluate chunks of input in context (i.e., deciding about the vulnerability of certain methods in relation to the presence of proper input checking), we decided to rely on a specific family of neural networks: \emph{recurrent neural networks} \cite{goodfellow2016deep}.
A RNN is a model that implements the following function:
\[
    \mathbf{s_n} = f (\mathbf{x_n}, \mathbf{s_{n-1}}),
\]
where $\mathbf{x_n}$ represents the $n^{th}$ input (or part of input), and $\mathbf{s_n}$ denotes the internal state of the RNN after processing the $n^{th}$ input. In our case, each input will represent a line of code in a program, and we will evaluate the last state $\mathbf{s_n}$ produced by the RNN to decide if the whole program contains a buffer overflow vulnerability.  

A long short-term memory (LSTM) \cite{hochreiter1997long} is a specific variant of a recurrent neural network which relies on gates to process and filter the information. This design choice allows LSTM to better model long-range dependence in the data. Such a feature is particularly important in our application as it allows to capture a wider context for potentially vulnerable instructions.

\section{Related work} \label{sec:RelatedWork}

In this section, we describe state-of-the-art work done on the problem of vulnerability detection, paying particular attention to types of data representation and model architecture adopted by other researchers. Although work on vulnerability detection has close connection to malware detection (see, for reference, recent work such as \cite{anderson2018ember} or \cite{raff2017malware}) and intrusion detection (refer, for instance, to \cite{bontemps2016collective}), we will focus mainly on vulnerability detection.

In \cite{hovsepyan2012software}, Hovsepyan et al. process Java source code treating it as plain text, and they train a SVM classifier to decide on the presence of vulnerabilities in the code. A similar approach is adopted in \cite{pang2015predicting}, where Java source code is represented via n-grams and synthetic features before being processed by a SVM trained to detect vulnerable programs. Representation learning via principal component analysis has been applied to C source code in order to generate informative representations for vulnerability detection in \cite{yamaguchi2011vulnerability}. All these approaches closely resemble our work in that they subscribe, however implicitly or partially, to the hypothesis that code can be dealt as natural language; however, our work relies on assembly code, making the reasonable assumption that source code is often unavailable, and it applies more versatile models, such as RNNs.

The use of CNN and RNN models has been considered in \cite{VDDR}, where Russel et al.  trained a vulnerability detection model on large data sets of real-world C/C++ source code. Similarly, in \cite{VDP}, a RNN relying on bi-directional LSTM, is shown to be more effective at vulnerability detection than other standard pattern-based systems such as Flawfinder, RATS, or Checkmarx.
Our approach resembles these studies in the choice of the family of models we consider (RNNs); however, once again, we decided not to consider source code, but, more realistically, assembly code.

A different direction of research has considered analyzing programs using not only static features, but also dynamic features, that is, feature generated by a program at runtime. Grieco et al. profiled code for vulnerability by using a collection of both static and dynamic features (such as program events), and by training different machine learning algorithms to learn useful patterns \cite{vdml}. Wu et al. developed a model relying only on dynamic features (such as kernel function calls) and implemented deep neural networks to carry out vulnerability detection \cite{vddl}.
Our work does not take into consideration dynamic features, although the architecture we have chosen would be versatile enough to be extended in the future to process such features.

\section{Problem definition} \label{sec:ProblemStatement}
In this work, we formulate and address the problem of discovering stack-based buffer overflow vulnerabilities in a program by processing its assembly code representation via RNNs. 

At its foundation, our work is grounded on the assumption that code constitute a form of language \cite{allamanis2018survey}. Code has a tightly structured syntax, and it can convey meaning in a similar fashion as written and spoken languages.  
Through our experiments, we assess the hypothesis that code may be processed as a form of language, and that this representation may be successfully employed to perform vulnearability detection. To do so, we aim at processing code in the form of text, with minimal pre-processing based on human prior knowledge, and using models that proved to be successful in dealing with natural language processing.

Concerning the specific form of language considered, we decided not to focus on one specific high-level programming language. Instead we considered code written in assembly language. Assembly language constitutes a formal language that provides a middle ground between human-friendly programming languages and machine instructions. On one side, programming languages are very succinct and interpretable, but they come with a wide variety of vocabulary and grammars; models for a given language may hardly be extended to other languages; moreover, in the real world, source code of a program may be rarely available. On the other hand, machine code is heavily hardware dependent and verbose, as it specifies in fine-grained details elementary operations to be carried out. 
The assembly language offers thus a language that strikes a reasonable balance between conciseness and availability. Conciseness is important both to reduce the cost of storing and processing the data; availability is relevant in order to have models that may be trained on a substantial amount of data and that can be deployed in the real world.

It is worth to underline that, so far, to the best of our knowledge, limited work has been done to exploit this specific level of representation for a program.

\section{Methodology and Implementation} \label{sec:Methodology}
In this section we discuss our methodological approach and our practical implementation: how we generated data for our problem, what sort of processing we performed on them, and finally we provide details about the model we implemented.

Overall, our method is based on the following steps: (i) generation of a library of safe and vulnerable functions, (ii) sampling and aggregation of functions into programs, (iii) compilation of programs into assembly files, (iv) compression of the assembly files by removing redundant information, (v) tokenization of the assembly files, (vi) partitioning of the samples in training and test data, (vii) training of the model. This approach provides us with complete control over every aspect of the simulations, from the data (size, complexity, and variability) to the model (architecture, depth, and other training hyper-parameters).

\subsection{Data generation}

Given the absence of standardized public data sets for stack-based buffer overflows, we first considered the challenge of assembling a suitable data set of vulnerabilities.

Given the availability of code online, a common approach to collect a data set of vulnerable code is to design a \emph{webscraper} to download C source code from open source projects, like the Linux kernel\footnote{\url{https://www.kernel.org/}}, or public repositories, such as Github\footnote{\url{https://github.com/}}. This approach would return large amounts of realistic data, although a good subset of the retrieved code may be umantained and of low quality (e.g., failing to compile). The main drawback is that all the data would be unlabelled. Labeling the data, so that it could be used for supervised learning in our RNN models, would be a non-trivial challenge; manual labelling would be extremely time-consuming, requiring several experts to analyze the code to understand if it has potential buffer overflows vulnerabilities or not; automatic labelling using existing tools does, in a way, defeat the purpose of our research as it would lead the RNN model to simply learn the function already encoded in the existing automatic tools. Even more critically, only a small percentage of the collected data would contain samples of stack-based buffer overflow vulnerabilities which would be relevant in our problem.

Therefore, we opted to create our own dataset. This approach has some clear advantages: (i) since we define each function and program, and we know exactly whether it is vulnerable to stack-based buffer overflow or not, labelling is automatic and unequivocal; (ii) we have full control over the code style used, allowing us to produce more or less heterogeneous samples; (iii) since the generation is automated, we can easily scale up the size of the data, thus making the fitting of complex functions feasible. 
On the other hand, we acknowledge the fact that custom-made datasets are often criticized as lacking in realism; while this point is true, we hold that our samples are representative enough to validate or to disprove our hypothesis that RNNs may be used for stack-based buffer overflow vulnerability detection. If we were to prove that our models can successfully detect unsafe code, more realistic data and deeper network may be implemented to tackle more realistic challenges. 



A \emph{sample} in our synthetic dataset is a C program file. A program file is made up of a variable number of \emph{functions} that are called in the main body of the program file. In the next section, we explain how we generate functions and how we assemble them into programs.

\subsubsection{Generating safe and vulnerable functions}
We view a function as the atomic component of a C program. Each function is built around a \emph{system call}. Although there are many potentially vulnerable system calls in the C library, we restrict our attention to $8$ system calls that are particularly relevant in the context of buffer overflows: \emph{strcpy, strncpy, strcat, scanf, sprintf, gets, fgets, memcpy}. Some of these system calls are deemed intrinsically unsafe and are deprecated, while other C methods are considered safe and are recommended. However, our labelling does not depend on this static distinction. We define a function non-vulnerable if it correctly uses a safe system call or if it employs an unsafe system call together with right buffer checks; vice versa, we define a function vulnerable if it uses an unsafe system call with wrong checks or if it misuses a safe system call. For instance, although \emph{strcpy} is normally considered unsafe, we recognize that it may be used safely when accompanied by proper buffer checks. By taking into account these more subtle differences, we aim at training a model that does not just trivially spot the use of deprecated system calls, but evaluates their vulnerability in context. 

For each of the $8$ system calls above, we created $15$ different functions with a single vulnerable system call. In total, this leads to $120$ different vulnerable functions. These functions have one and only one vulnerability and use one and only one weak system call, thus defining simple functions with a single vulnerability, while preserving the realism of our data.

For the benign functions, we decided to create an equal amount of instances. Among these instance, we included functions properly using safe system calls as well as repaired versions of the unsafe functions used to create vulnerable functions. Listing \ref{notSafeMemcpy} presents a vulnerable function along with its safe counterpart in Listing \ref{safeMemcpy}. The two listings show how easy it is to unintentionally introduce a vulnerability by using the wrong parameter in a C system call, and how subtle the difference between a safe and an unsafe function can be. These minimal differences allow us to verify whether a network is able to discriminate safe and unsafe programs given the context in which C function calls happen, and not just by the name of the function called.

\begin{lstlisting}[backgroundcolor=\color{backgroundColour}, label=notSafeMemcpy, caption= A vulnerable function]
void memcpySmallIntoLarge(char* s){
    char dest[256];
    memcpy(dest,s,strlen(s));
    printf("%s\n", dest);
}
\end{lstlisting}

\begin{lstlisting}[backgroundcolor=\color{backgroundColour}, label=safeMemcpy, caption= Repaired version of Listing \ref{notSafeMemcpy}]
void memcpySmallIntoLarge(char* s){
    char dest[256];
    memcpy(dest,s,sizeof(dest));
    printf("%s\n", dest);
}
\end{lstlisting}

\subsubsection{Generating C programs}

From the set of functions we have defined, we generate programs. We define \emph{positive sample} a C program that contains no vulnerable function; conversely, we define \emph{negative sample} a C program that contains at least one vulnerable function. We acknowledge that functions normally are vulnerable through their calling parameters and the state of the virtual address space. However, we limit ourselves to define vulnerable and benign samples without additional considerations such as those aforementioned.

Given a number $N_f$ of function to be included in a program, we construct positive samples by randomly sampling $N_f$ safe functions, inserting their definition into a C program, and adding function calls in the \emph{main()} method; for negative samples, we randomly select a single unsafe function along with $N_f-1$ random safe functions, and we aggregate them in a C program as done for positive samples.
The resulting programs are carefully checked to verify whether they are actually safe or unsafe. We followed a unit testing approach, validating each unit of code through a range of inputs.


\subsection{Data processing}

The output of our data generating process is a set of programs in the form of C source code. Since we do not want to work at this level of representation, we compile our data into assembly. Moreover, before feeding the data to the model, we further process it by removing redundant information, converting it into token strings, and finally partitioning it into training and test data.

\subsubsection{Compiling C programs into assembly}

First, we translate our source code into assembly code, using a compiler with Intel syntax for the 64-bit architecture.
Compilation happens using the GCC compiler \cite{GCCWiki}, a versatile and cross-platform compiler widely adopted on many systems. Our C programs are compiled running the following command:
\begin{lstlisting}[style=argvListing]
gcc -S -fno-asynchronous-unwind-tables -masm=intel ./fileName.c -o fileName 
\end{lstlisting}
For the sake of generalization, we rely on a minimal number of flags aimed at producing as short an assembly as possible; for details about the flags, please refer to \cite{GCCWiki}. A snippet of generated assembly is shown in Listing \ref{assembly}.

\begin{lstlisting}[style=customasm, label=assembly, caption= Example of part of a C program compiled in assembly]
.file "test_file_3_0.c"
.intel_syntax_noprefix
.text
.glob main 
.type main, @function

.LC0:
    .string "Enter the size of input:"
.LC1:
    .string "%d"
    
mov    rbx, rax
lea    rdi, .LC0[rip]
mov    eax, 0
call   printf@PLT

push    rbx
sub     rsp, 72
mov     QWORD PTR -104[rbp], rdi

\end{lstlisting}

This processing step makes our model independent from the availability of C source code. Real-world executables, for which the source code has not been released, can still be decompiled into assembly language. Several tools, such as IDA Pro by Hex Rays \cite{idaPro}, Binary Ninja \cite{binaryNinja}, and NSA's own Ghidra \cite{ghidra}, allow for decompiling binaries into assembly code. On the other hand, succesfully reverse engineering machine code up to C is a more challenging, and still open, problem.
Thus, processing assembly code instead of C codes makes our model more flexible and generic, potentially able to deal with a larger class of real-world samples.



\subsubsection{Compressing the assembly code}

Next, in order to simplify the model and make the learning process faster, we compress the assembly code by discarding redundant information.

Several lines of the assembly code generated by the GCC compiler have limited value, containing either redundant or useless information for our aims. 
We then start by removing the entire prefix of each assembly (see lines 1-5 in Listing \ref{assembly} for an example). The prefix is equal for each assembly except the field \emph{.filename} which just reports the name of the original file; as such, no significant information is carried in these lines of code.


Next, we consider other assembler directives and instructions that can be dropped \cite{assDirectives}:
\begin{itemize}
    \item \emph{.LC} directives are used for storing declared strings in the C program (see line 7-10 in Listing \ref{assembly} for an example). The actual content of a string is not relevant in our modelling. However, in order to maintain as much as possible of the code context, notice that we keep references to the strings (see line 13 in Listing \ref{assembly} for an example).
    \item \emph{.size} directives are generated by compilers to include auxiliary debugging information in the symbol table. Compiler information is not relevant in our modelling.
    \item \emph{.ident} directives are used by some assemblers to place tags in object files. Object generation is not relevant in our modelling.
    \item \emph{.section} directives are used to manage sections in objects. 
    Object management is not relevant in our modelling.
    \item \emph{endbr} instructions \cite{intelSpec} in the last part of each assembly are removed.
\end{itemize}

The end result is shorter assembly samples that still preserve the semantics of the code. Notice that in this processing steps we minimally rely on prior knowledge. An expert evaluation is indeed needed to decide what parts of an assembly code to discard. However, we argue that such choice boils down to a non-case-specific and procedural rule that consistently drop pre-determined lines. This pre-processing does not introduce any form of high-level knowledge, nor does it instantiate any sort of feature carrying human-injected knowledge. The point of this pre-processing is simply to reduce the size of the samples in order to limit the computational burden. It is our conjecture that, if our model were to be successful, scaled-up models would be able to successfully process non-compressed versions of the same assembly.


\subsubsection{Tokenization of the assembly code} 
Following the standard in natural language processsing, we proceed with a tokenization of the assembly code. Although the main component of an assembly may be identified by a line of code, lines present too much variety to be reduced to tokens. Instead we split lines of code into atomic codewords, and we use regular expressions to meaningfully extract tokens. Regular expressions are designed to preserve relevant commands and their context, while at the same time lead to a restricted dictionary of meaningful symbols. As an example, consider line 19 in Listing \ref{assembly}; its tokenization returns: 
\begin{lstlisting}[style=argvListing]
    "mov", "QWORD", "PTR", "-104", "[", "]", ",", "rbp", "rdi", "/n"
\end{lstlisting}
Notice that in our tokenization we always keep the newline character '/n' to denote the end of lines of code, as we consider information on the end and the beginning of a line relevant in our analysis.

\subsubsection{Partitioning of the data}
At the end of this process, we are left with a data set where each sample, positive or negative, is a set of assembly tokens. According to the good practices of machine learning, we split this data set in a training data set (80\% of the data) and a test data set (20\% of the data). Training data are further partitioned in a set used for training our model and a development (or validation) set for hyperparameter optimization. Test data are used only for assessing the accuracy of our model.

\subsection{Model definition}

In order to assess our hypothesis, we adopted a standard neural network model widely used in NLP and we applied it to the processing and classification of assembly code. 
Given the available computational limitation, we implemented an effective, yet lightweight, RNN architecture designed to discriminate vulnerable and non-vulnerable samples taking into account both the local and the global context in each program.

\begin{figure}[H]
    \includegraphics[scale=0.3]{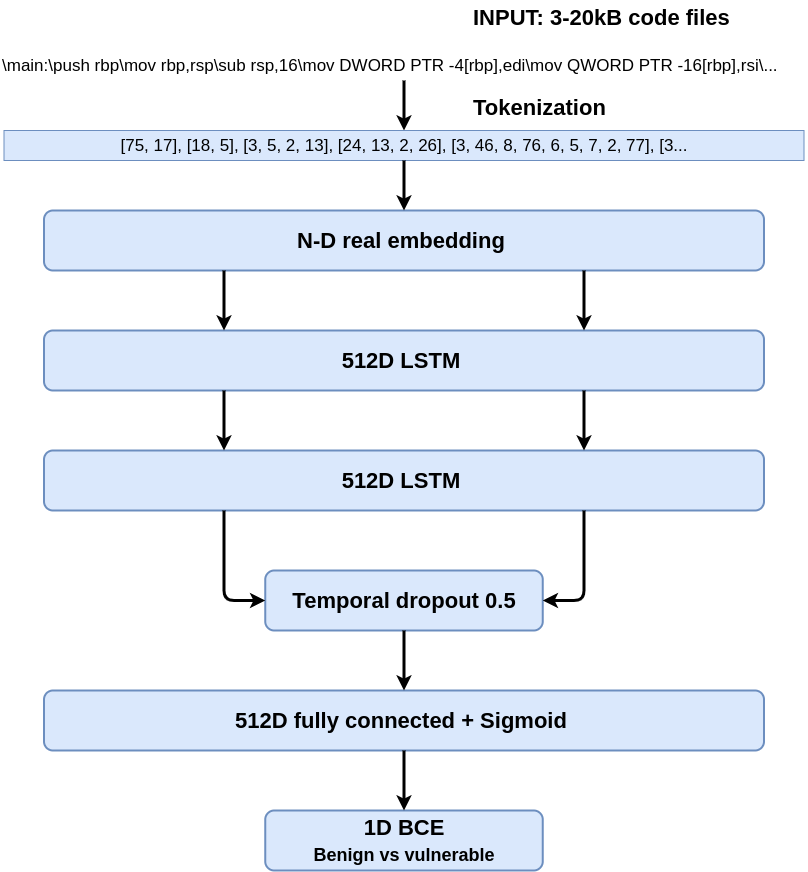}
    \caption{Architecture of our model}
    \label{fig:architecture}
\end{figure}

With reference to Figure \ref{fig:architecture}, our architecture is designed as follows. The model receives as input a tokenized assembly file. Each token is mapped to an integer value via a pre-computed table, thus creating an integer array. This vector is forwarded to a 512-dimensional embedding layer which maps the discrete representations into dense representations. Two hidden LSTM layers follow, with the task of processing the data and tracking the inner state of the network; temporal dropout is applied to the last layer of these two layers in order to improve the generalization performance. Finally, the output of the LSTM layers is forwarded to a fully-connected sigmoidal layer that produces the decision of the model.
The entire model is trained using a binary cross-entropy loss \cite{goodfellow2016deep}.

\section{Experiments} \label{sec:Experiments}

In this section we present and discuss the simulations we have done in order to prove or disprove the language hypothesis applied to assembly code and to verify the feasibility of detecting stack-based buffer overflow using RNNs.

Our RNN models have been implemented using PyTorch \cite{NIPS2019_9015}, an open-source machine learning library for Python. The source code for our model is available publicly online at \url{https://github.com/williamadahl/RNN-for-Vulnerability-Detection}.

In all the simulations, we use a training set to train the model, a development (or validation) set for model selection, and a test set to evaluate the final performance. We track the loss on training and development set during training, and we measure the final performance in terms of correct classification rate (CCR, or accuracy).

\subsection{Simulation 1: Detecting vulnerabilities}

In this first simulation, we evaluate the network ability to discriminate between programs composed by safe functions and programs containing different vulnerable functions. Positive and negative samples use clearly different functions, thus presenting to the network a relatively easy discrimination task.


\subsubsection{Protocol.}

In our first simulation we sampled $2000$ benign and $2000$ vulnerable binaries. Each binary contains $3$ functions where the benign functions are sampled from the set of safe functions excluding repaired versions of vulnerable functions, while the vulnerable functions are sampled from the whole set of vulnerable functions. 

Training and development were conducted considering the batch sizes $\{20,40,80, 100\}$, while the learning rate was set relatively aggressive considering the values in the set $\{0.00025, 0.0005, 0.001, 0.002\}$.

\subsubsection{Results.}

\begin{figure}[H]
    \includegraphics[scale=0.42]{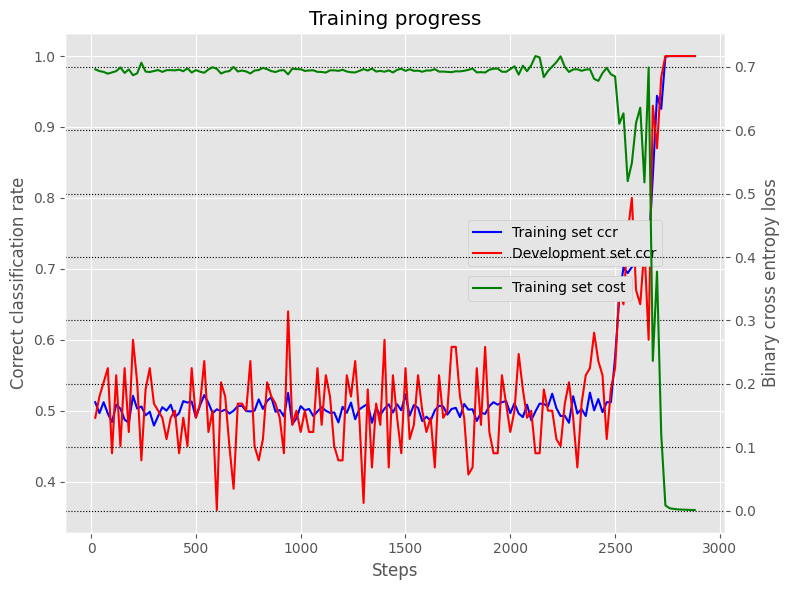}
    \caption{Performance for Simulation 1 using learning rate $10^{-3}$ and a batch size of $80$.}
    \label{fig:ex1_plt0}
\end{figure}

We selected the best configuration of hyperparameters (batch size $80$, and learning rate $10^{-3}$) in terms of performance on the development data set; other configurations achieved lower results, and often required more epochs to achieve convergence.

Figure \ref{fig:ex1_plt0} shows the result of training. Notice the two scales for the $y$-axis: on the left we report the correct classification rate (CCR) for the training (blue) and the development (red) dataset; on the right we report the loss function for the training dataset (green). At the end of training the final performance on the test set successfully converges to a CCR of $1.0$.

Even with a modest dataset of 4000 samples we can observe in Figure \ref{fig:ex1_plt0} that we were able to achieve a very low loss of 0.001 and a CCR of 1.0 over the training \emph{and} development set. We can deduce from this experiment that the neural network exhibits the capacity to classify with very low error over the subset of functions we defined. 


\subsubsection{Discussion.}
In this simulation, the model we designed proved able to discriminate with high accuracy between safe programs and vulnerable programs. However, what we considered is just a simplified problem, where positive and negative samples are clearly different. It may be hypothesized, then, that the RNN modeled generic patterns in our samples, which may not be tightly related to buffer overflow. 


\subsection{Simulation 2: Differentiation between vulnerable and repaired counterparts}

To disprove the hypothesis that the RNN is not meaningfully capturing buffer overflow vulnerabilities, in this simulation we consider a more challenging scenario, in which vulnerable functions appear in negative samples, while the fixed version of the same vulnerable functions appears in positive samples. Vulnerable functions and their patched counterparts are very similar in code, often with only one line differentiating a positive from a negative sample. This setup allows us to probe thoroughly the discriminative power of the network. 


\subsubsection{Protocol.}

We sample negative samples as functions containing a vulnerability from a single type of system-call, \emph{fgets}. Positive samples are built including the patched counterparts of the vulnerable functions; these patched functions still include the call to \emph{fgets}, but in a safe and controlled way. In total, we train the network with $200$ samples of both classes.

We experimented with low batch sizes in the set $\{2, 5, 10\}$, due to the relative small sampled dataset, and learning rates in the set $\{0.000125, 0.00025, 0.0005, 0.001, 0.002\}$.

\subsubsection{Results.}

\begin{figure}[H]
    \includegraphics[scale=0.42]{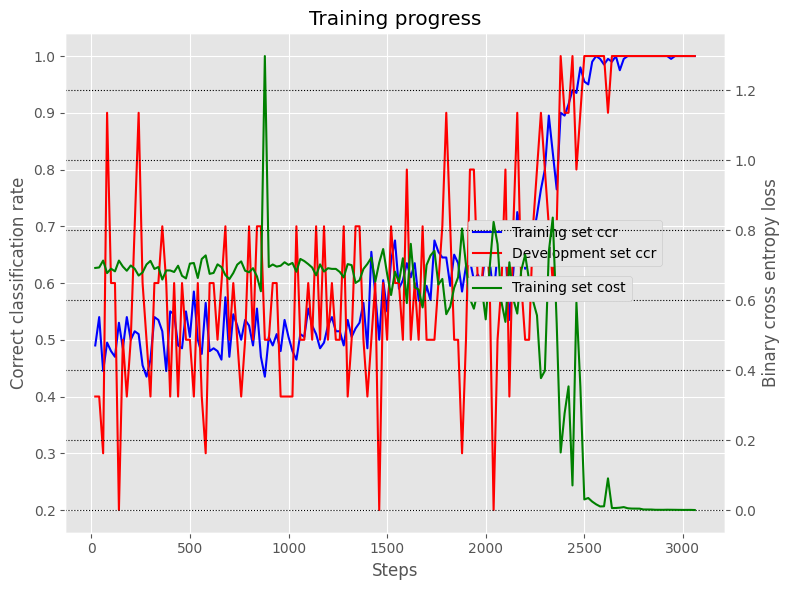}
    \caption{Performance for Simulation 2 using learning rate $1.25 \cdot 10^{-5}$ and a batch size of $10$.}
    \label{fig:ex2_plt0}
\end{figure}

The best hyperparameter configuration we found uses a batch size of $10$ and a learning rate of $1.25\cdot{10}^{-4}$. Figure \ref{fig:ex2_plt0} shows the dynamics of training, with the model achieving again a CCR close to $1.00$ on the training \emph{and} development set. CCR on the test set was similarly close to 1.0. 


Different choices for the hyperparameters, such as a learning rate above $2.5\cdot{10}^{-3}$ often got stuck and failed at learning. We hypothesize that the high similarity among positive and negative samples combined with a relatively high learning rate may lead to an oscillatory behaviour, where the network tends to alternatively overshoot or undershoot with respect to the thin margin separating the classes. 
Models trained with lower learning rate achieved overall better performances, although in some cases, even models with a learning rate of $5.0\cdot{10}^{-3}$ could achieve a satisfying CCR, despite showing high instability on average. 


\subsubsection{Discussion.}
Our model was able to successfully learn to discriminate between subtly different samples. This confirm that the RNN model is able to capture subtle and contextual features of the assembly code that actually relate to buffer overflow vulnerabilities. This result is particularly relevant because, in the real world, the difference between a safe program and a vulnerable executable may be as small as a single line of code; performing a length check on an input, or size check for the allocation of a large buffer, may distinguish a safe program from an unsafe one. The high accuracy achieved by our network suggests that machine learning model may be successfully deployed to capture subtle flaws in code. 


\subsection{Simulation 3: Larger datasets and samples}

In this last experiment, we test our model on a larger dataset drawn from an even more extensive range of functions. We wish to push our model further and simulate more realistic code samples.

\subsubsection{Protocol.}

In this final experiment we consider a collection of $4000$ benign and $4000$ vulnerable samples. Each sample consist of 20 functions, where the benign functions in both positive and negative samples are drawn from the same distribution. This distribution does not contain repaired counterparts to the vulnerable functions in the vulnerability library. The negative samples are drawn at random from the whole library of vulnerable functions. 

We trained a model made up by a two layer LSTM architecture with a fixed batch sizes of $80$, and a learning rate in the set
$\{0.0001, 0.00025, 0.0005, 0.001\}$. We chose these hyperparameter ranges as they have proven successful in previous experiments on larger datasets.

\subsubsection{Results.}

\begin{figure}[H]
    \includegraphics[scale=0.46]{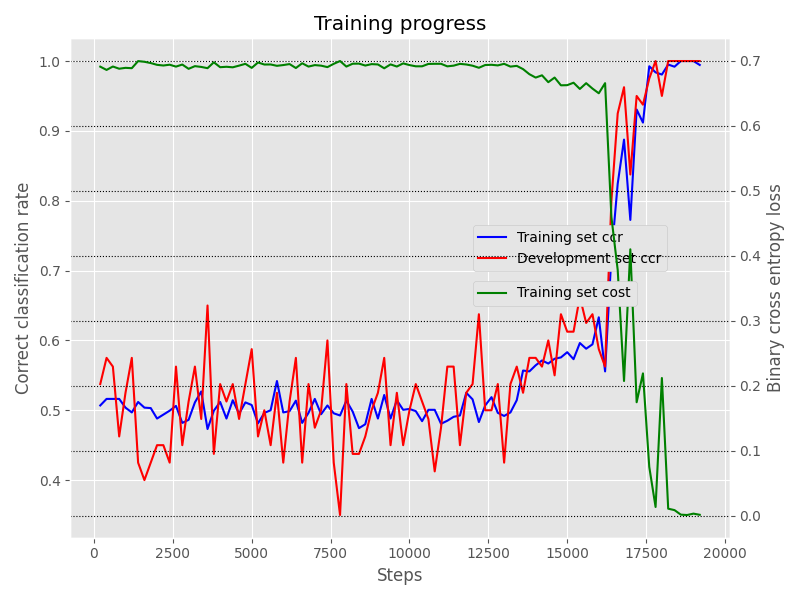}
    \caption{Performance for Simulation 2 using learning rate $5\cdot{10}^{-5}$ and a batch size of $80$.}
    \label{fig:ex3_plt0}
\end{figure}

Figure \ref{fig:ex3_plt0} shows the results achieved for the two layers LSTM configuration. From the graph, we observe an improvement in the long-term in terms of CCR on the training set and the development set. 
After about 17,500 steps, the neural network breaks through a plateau, and drastically improves CCR and reduce loss over the training and the development set. At termination our model achieved a CCR of 99.00\% and a 0.057 loss over the test set.
The neural network was capable of processing and learning features for samples of substantial size, and converge within a reasonable time frame. These observations are encouraging as the "real world" code encompass a wide range of sizes and structure. 

For this particular experiment we also set up a shallower architecture with only one layer of LSTM. This simplified architecture did not perform as well as the two layer LSTM within a comparable time frame.

\subsubsection{Discussion.}
Although quite sensitive to hyperparameter tuning, the RNN model with the right setting (a batch size of 80 samples was by far the most successful configuration) yielded a model that achieved a satisfactory result within the training time constraint. It is interesting to observe that we had to apply a relatively low learning rates throughout our experiment to achieve satisfactory results. 
In conclusion, we can state that our neural network proved able to learn from complex samples made up of multiple functions.

\section{Discussion} \label{sec:Discussion}

Our simulations allows us to conclude that not very deep RNNs are able to successfully perform binary classification on the problem of detecting the presence of stack-based buffer overflow vulnerabilities. Even when presented with samples where discriminating between a positive and negative instance relied heavily on the context, the trained model was able to differentiate satisfactorily.

The results provide a proof-of-concept validation of the language hypothesis we presented in Section \ref{sec:ProblemStatement}. However, our conclusions are limited by the relative simplicity of our data sets and our models.
Our self-generated data set can hardly be used to generalize over the large variety of real-world data. Overcoming this limitation would require either the collection and the labeling of large quantities of code samples, or the refinement of generators used to create realistic code samples. In this last case, our own generator may be improved by introducing additional functions, more complex code structure, and recursive calls. Scaling up the complexity of the data has to be accompanied by a similar scaling in the depth and complexity of the RNN model. This would put more pressure on computational resources, as longer training and more hyperparameter tuning may be in order to produce a satisfactory model.

\section{Conclusion} \label{sec:Conclusion}

Our research aimed at exploring the application of the language hypothesis to assembly code, and more specifically, at understanding the possibilities and limitations of stack-based buffer overflow vulnerability detection using RNNs. In order to prove or disprove the hypothesis, we conducted a set of experiments on binary classification. Our results showed that RNNs are able to extract useful features from the assembly code language and evaluate the context of the instructions. Such conclusions lend support to the hypothesis on the similarity between natural language and programming languages, and provide ground for treating code as language in future research. 

Several possible extensions for further development lie open ahead. The easiest direction would be to scale up our model in terms of network depth, dataset size and realism; this would require a more significant computational effort, but it may further confirm (or, possibly, highlight the limitations of) the language hypothesis. Another direction would be to focus on alternative model architectures: we only considered LSTM layers, but layers with attention mechanisms \cite{attmek} may provide ground-breaking results, as they have done in standard NLP tasks \cite{attNeed}. Our model could also be enriched to perform multiclass prediction (thus distinguishing different types of buffer overflows) or to output not only the presence or absence of a vulnerability, but also its location (thus allowing to identify and fix buffer overflows more easily).
Finally, our approach may be applied to other types of vulnerabilities; buffer errors have been the most ubiquitous type of vulnerability for the last 25 years \cite{topIncident25year}, but other vulnerabilities, such as race conditions or failures to validate input, can be equally harmful and may be predicted with similar RNN models.

\ifCLASSOPTIONcaptionsoff
  \newpage
\fi



\bibliographystyle{IEEEtran}
\bibliography{IEEEabrv,bibliography}
%



%



\begin{IEEEbiographynophoto}{William~Arild~Dahl}
graduated with a master's degree in informatics at the University of Oslo. His research topic was machine learning for vulnerability detection. Currently he is working as a DevOps engineer at Netcompany Norway.  

\end{IEEEbiographynophoto}

\begin{IEEEbiographynophoto}{L{\'a}szl{\'o}~Erd{\H o}di}
has a PhD in cyber security. Currently he is a lecturer and researcher in the information security research group at the University of Oslo. His main research field is offensive security. He is the leader of the Hacking Arena.
\end{IEEEbiographynophoto}

\begin{IEEEbiographynophoto}{Fabio Massimo Zennaro}
received his PhD in Machine Learning from the University of Manchester. Currently he is a postdoc researcher in the Digital Security and in the Oslo Analytics research groups at the University of Oslo. His research interests include representation learning, causality, fairness, and security.
\end{IEEEbiographynophoto}






\end{document}